\documentclass[conference]{IEEEtran}
\usepackage{cite}

\ifCLASSINFOpdf
\else
\fi

\usepackage{graphicx}
\usepackage{xcolor}
\usepackage{mathrsfs}
\usepackage{amssymb}
\usepackage{amsmath}
\usepackage{bbm}
\usepackage{algorithm}
\usepackage{algpseudocode}
\usepackage{gensymb}
\usepackage{mathtools}
\usepackage{fancyhdr}

\begin{document}
%
\title{Economic Battery Storage Dispatch \\with Deep Reinforcement Learning from Rule-Based Demonstrations}

\author{\IEEEauthorblockN{Manuel Sage}
\IEEEauthorblockA{Department of Mechanical Engineering\\
McGill University\\ Montreal, Canada\\
manuel.sage@mail.mcgill.ca}
\and
\IEEEauthorblockN{Martin Staniszewski}
\IEEEauthorblockA{Siemens Energy Canada Limited\\
Montreal, Canada\\
martin.staniszewski@siemens-energy.com}
\and
\IEEEauthorblockN{Yaoyao Fiona Zhao}
\IEEEauthorblockA{Department of Mechanical Engineering\\
McGill University\\ Montreal, Canada\\
yaoyao.zhao@mcgill.ca}}


%


\maketitle
\pagestyle{plain}  
\thispagestyle{fancy}  
\fancyhf{} 
\renewcommand{\headrulewidth}{0pt} 
\cfoot{Published as a conference paper at ICCAD 2023}  

\begin{abstract}
The application of deep reinforcement learning algorithms to economic battery dispatch problems has significantly increased recently. However, optimizing battery dispatch over long horizons can be challenging due to delayed rewards. In our experiments we observe poor performance of popular actor-critic algorithms when trained on yearly episodes with hourly resolution. To address this, we propose an approach extending soft actor-critic (SAC) with learning from demonstrations. The special feature of our approach is that, due to the absence of expert demonstrations, the demonstration data is generated through simple, rule-based policies. We conduct a case study on a grid-connected microgrid and use if-then-else statements based on the wholesale price of electricity to collect demonstrations. These are stored in a separate replay buffer and sampled with linearly decaying probability along with the agent’s own experiences. Despite these minimal modifications and the imperfections in the demonstration data, the results show a drastic performance improvement regarding both sample efficiency and final rewards. We further show that the proposed method reliably outperforms the demonstrator and is robust to the choice of rule, as long as the rule is sufficient to guide early training into the right direction. 

\end{abstract}


%

\section{Introduction}
Energy storages play a vital role in the transition towards “green” electricity grids. By smoothening the intermittent power production of renewable energies (RE) and by shaving peaks in electricity demand, storages facilitate the large-scale integration of renewables. Batteries, in particular lithium-ion batteries, have emerged as one of the most promising storage technologies due to significant technological progress and cost reduction \cite{a201_ziegler_battery_cost}. This has led to the planning and realization of utility-scale projects with previously unimaginable capacities. A popular example here is Vistra’s Moss Landing Energy Storage facility in California, United States, consisting of batteries with 400 MW power and 1,600 MWh energy capacity that are currently expanded to 750 MW/ 3,000 MWh \cite{w20_vistra_battery_project}.

For the proper planning of such facilities, control algorithms are required that can model plant operation in uncertain conditions, for example regarding electricity prices and RE production, and optimize the dispatch of batteries. The objective thereby is usually either minimizing operational cost or maximizing profits and the results can help to dimension the system and to choose appropriate components and locations. In the past few years, reinforcement learning (RL) has gained attention as a tool for economic battery dispatch and research in this field has increased substantially \cite{a194_subramanya9777914}.

Especially deep reinforcement learning (DRL) algorithms with actor-critic architectures such as proximal policy optimization (PPO) \cite{a121_schulmanPPO} and deep deterministic policy gradients (DDPG) \cite{a123_lillicrap2019continuous} have thereby performed well. For example, Zhang et al. \cite{a91_zhang2021data} model a hybrid plant consisting of wind, solar photovoltaic (PV), diesel generators, batteries, and reverse osmosis elements producing power and freshwater. Task of the employed control algorithms is to dispatch the diesel generators and batteries efficiently and minimize system cost. Three popular DRL algorithms, namely double deep Q networks (DDQN), DDPG, and PPO, are compared with the non-RL alternatives stochastic programming (SP) and particle swarm optimization. The temporal resolution of the task is 1h and an episode is 24h long. PPO was found to perform best and to achieve a cost reduction of over 14\% compared to SP.

DDPG has been compared to model predictive control (MPC) in the work of da Silva Andr\'e et al. \cite{a193_DASILVAANDRE2022108551} on battery dispatch for two different electric grid models. The duration between timesteps is 9s with 100 timesteps per episode. Despite not having access to forecasts, contrary to MPC, DDPG achieves similar cost reduction while significantly reducing the computational time. Zha et al. \cite{a195_zha2021improved} propose a modified actor-critic architecture with a distributional critic network to optimize a grid-connected solar PV, wind, and battery hybrid system. The modifications include grouping four subsequent observations into a state and then using convolutional neural networks for function approximation as well as the pretraining of the critic for 200 episodes. The researchers train the algorithm using an hourly resolution and 720 timesteps (one month) per episode and report an increased performance of this variant over the tested alternatives. The wide-spread applications of reinforcement learning to battery control have been reviewed in detail by Subramanya et al. \cite{a194_subramanya9777914}.

The main advantage of popular RL algorithms for battery control is their model-free optimization. Instead of requiring a full mathematical model of the investigated environment, only a sample model is needed \cite{a178_li2020}. This leads to a data-based approach to optimization that, besides simplifying implementations, allows to capture uncertainty. Modern power systems are complex in nature and subject to different types of uncertainty, for example regarding demands and renewable power production. Model-free RL can cope with this challenge by optimizing complex systems over long periods \cite{a162_perera2021applications, a176_zhang2020, a207_yang2020reinforcement}.
On the other hand, RL often struggles to find good policies on tasks with sparse or delayed rewards \cite{a208_RUDDER}. As the purpose of batteries is to store energy for later use, delayed rewards are a natural phenomenon. Moreover, agents must learn that investing energy by charging the battery is a requirement for receiving these rewards at later timesteps.

In our experiments, we observe that DRL algorithms struggle to learn useful policies across various algorithms, implementation choices, and environment configurations. Often, training does not exceed the performance of random agents or collapses to a single action, leading to de-facto ignoring the battery. By modeling entire years in 8760 hourly timesteps, our episodes are significantly longer compared to the related work referenced above. This aggravates the dispatch task due to larger state spaces and potentially longer intervals between charging and discharging cycles. However, longer modeling periods capture more uncertainty and lead to more expressive results. In order to obtain good results while keeping the episodes long, we resort to learning from demonstrations, a branch of RL that has previously shown to boost performance during the initial stage of training.

The idea that learning from demonstration, for example from human experts or previous control systems, facilitates learning problems in RL is well-established \cite{a202_schaal_LfD, a199_hester2018dqfd, a192_reddy2019sqil, a200_gao2018reinforcement}. One of the most popular applications to DRL has been deep Q-learning from demonstrations (DQfD), an approach that achieved state-of-the-art results and sample efficiencies on many of the tested Atari games when introduced by Hester et al. \cite{a199_hester2018dqfd}. The researchers first pretrain the Q-network on demonstrations from a human player using an additional supervised loss term. Then, the agent starts to interact with the environment, storing its experiences in the same replay buffer but never overwriting the demonstrations. Sampling from the replay buffer occurs with prioritized experience replay \cite{a196_schaul2015prioritized}. The demonstrations are given an extra priority bonus to increase their sampling frequency.

Most approaches assume (near-) optimal demonstrations and force the agent to mimic the behavior of the provided examples. This is achieved either by additional supervised loss-terms as in DQfD, or by modifying the reward function to reward the agent only for behavior resembling the demonstrations. The latter is called imitation learning and has been applied to battery dispatched by Krishnamoorthy et al. \cite{a191_krish}. The researchers deploy soft actor-critic (SAC) with imitation learning to control batteries in 34-bus and 123-bus systems over 450 timesteps with 10ms resolution. The demonstrations stem from previously solving the same system using mixed-integer linear programming (MILP) and are assigned a constant reward of +1, while experiences from agent-environment interactions receive a reward of 0. Both types of data are sampled with equal probability during training. This approach, originally proposed by Reddy et al. \cite{a192_reddy2019sqil}, greatly reduces the computational time compared to MILP, but it can only be applied in settings where perfect demonstrations exist. Gao et al.
\cite{a200_gao2018reinforcement} demonstrate that RL can benefit from imperfect demonstrations as well. The researchers propose normalized actor-critic (NAC), an algorithm that does not conduct supervised pretraining and instead uses an additional normalization term for the gradient of the loss function to reduce the bias from demonstration data. The experiments on driving games show that NAC surpasses the demonstrator and other algorithms, even when fed with imperfect demonstrations.

In this study, we propose an approach to improve economic battery dispatch with RL through imperfect, rule-based demonstrations. We collect demonstrations by controlling the same environment via trivial if-then-else statements based on electricity prices. Our approach uses SAC \cite{a204_haarnoja2018sac2} and only requires minimal changes to the algorithm, namely a second replay buffer for demonstrations and a linearly decaying sampling ratio between the two buffers that favors the demonstrations initially. We show empirically with a case study that despite the simplicity, the approach is robust to the choice of rules and able to outperform both the demonstrator as well as other DRL algorithms and traditional methods. To the best of our knowledge, this is the first application of both imperfect and rule-based demonstration to battery dispatch with RL.

The remainder of the paper is organized as follows. The next section formulates the problem of economic battery dispatch. Section 3 introduces the modified SAC algorithm. Section 4 describes and discusses the conducted experiments before section 5 provides our conclusion.

\section{Problem Formulation}\label{problem}

\begin{figure}
	\begin{center}
		\includegraphics[width=8.4cm]{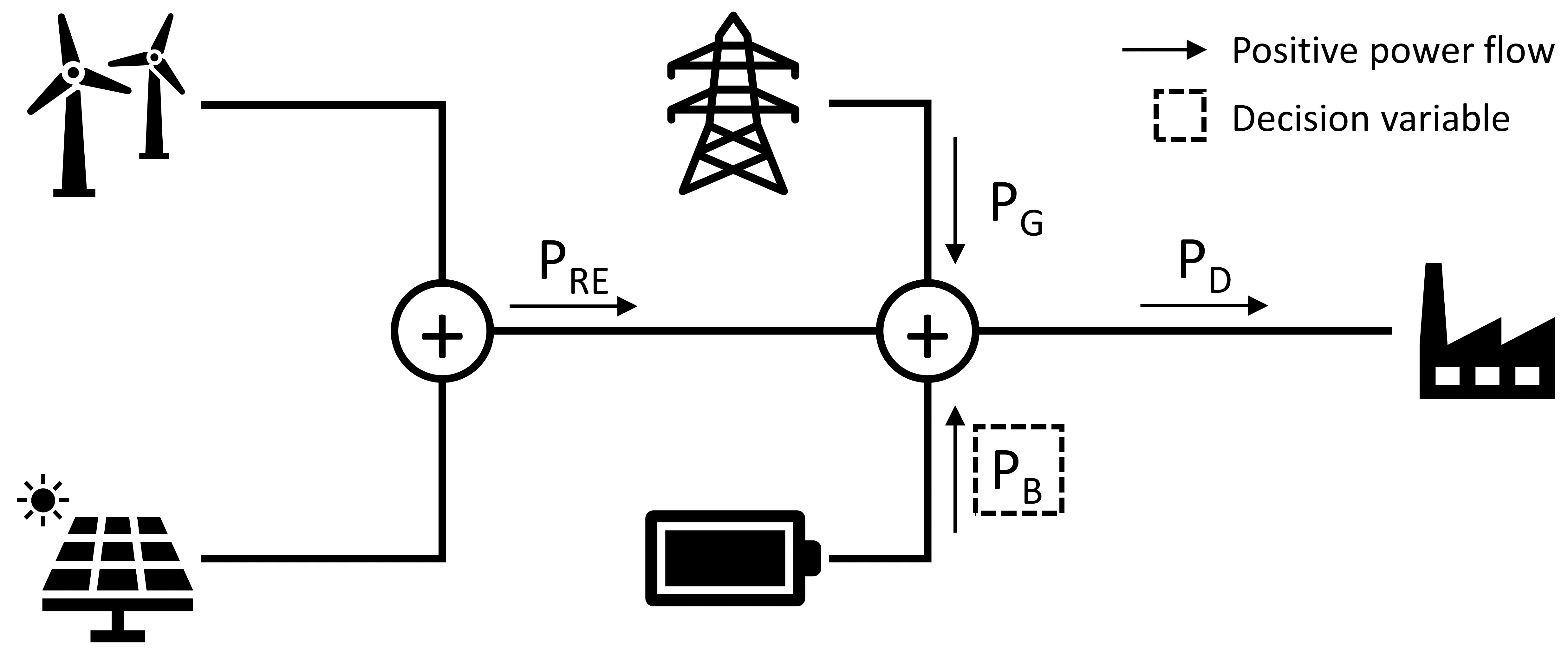}    
		\caption{Simplified model of the grid-connected microgrid.} 
		\label{fig:system}
	\end{center}
\end{figure}

Figure \ref{fig:system} shows the key components and power flows of the modeled hybrid energy system. We assume that the microgrid is connected to the utility grid and supplies its power to a factory with varying demands. Besides, it consists of wind turbines, solar PV modules, and the battery storage.

The objective is to minimize the total cost of supplying power to the factory, $C_{Total}$: 

\begin{equation} \label{eq1}
	\min C_{Total} = \sum_{t=0}^{T}C_{G, t}
\end{equation}
subject to
%
\begin{equation} \label{eq1b}
	C_{G, t} = 
	\begin{cases}
		P_{G, t} \times (c_{w,t} + c_{a}) & , P_{G, t} \geq 0\\
		P_{G, t} \times c_{w,t} & , P_{G, t} < 0
	\end{cases}
\end{equation}
\begin{equation} \label{eq2}
	P_{D, t} = P_{RE, t} + P_{B, t} + P_{G, t}
\end{equation}
\begin{equation} \label{eq3}
	SOC_{t} = SOC_{t-1}(1-\sigma) - \eta \frac{P_{B, t}\Delta t}{Ca V}
\end{equation}
\begin{equation} \label{eq4}
	SOC^{min} \leq SOC_{t} \leq SOC^{max}
\end{equation}
\begin{equation} \label{eq5}
	P^{min}_{B} \leq P_{B,t} \leq P^{max}_{B}
\end{equation}
\begin{equation} \label{eq6}
	P_{B,t} \geq -P_{RE, t}
\end{equation}

\noindent
$t$ = timestep\\
$C_{G, t}$ = cost of interacting with the utility grid\\
$P_{G, t}$ = power exchange with utility grid\\
$c_{w, t}$ = wholesale price for power\\
$c_{a}$ = fixed, auxiliary cost for purchasing power\\
$P_{D, t}$ = load/power demand\\
$P_{RE, t}$ = available power from renewables\\
$P_{B, t}$ = battery charging or discharging power\\
$SOC_{t}$ = state of charge (SOC) of battery\\
$SOC^{min/max}$ = minimum/maximum allowable SOC\\
$P^{min}_{B} / P^{max}_{B}$ = maximum battery charge/discharge rates\\
$\sigma$ = battery self-discharge rate\\
$\eta$ = battery charging and discharging efficiency\\
$CaV$ = capacity of the battery\\

In eq. \ref{eq1}, we assume that the operational cost of all components are fixed and thus not relevant for the optimization problem. This leaves the cost of interacting with the utility grid, $C_{G, t}$, as main cost factor. $P_{G, t}$ has a positive sign if power is bought from the grid and a negative sign if it is sold to the grid. Surplus electricity is sold at wholesale price, $c_{w, t}$, while purchasing deficit power adds fixed auxiliary cost, $c_{a}$, for example for transmission charges (see eq. \ref{eq1b}).

Constraint \ref{eq2} ensures that the power demand is met at all times. $P_{B, t}$ is negative if the battery is charged, and positive if it is discharged. Constraint \ref{eq3} describes the behavior of the battery's state-of-charge (SOC), with $\eta < 1$ during charging ($P_{B, t} < 0$) and $\eta > 1$ during discharging. Constraint \ref{eq4} restricts the permitted range of SOCs to prevent damage to the battery. Constraint \ref{eq5} limits $P_B$ to the maximum charging and discharging rates. We further add constraint \ref{eq6} to limit the battery charging to the available RE power.

\section{Proposed Method}\label{method}
To apply reinforcement learning to the battery dispatch problem, we resort to the common Markov Decision Process (MDP) framework \cite{a66_sutton2018reinforcement}. The MDP is defined as the tuple $(\mathscr{S}, \mathscr{A}, \mathscr{P}, \mathscr{R}, \gamma)$, consisting of the state space $\mathscr{S}$, the action space $\mathscr{A}$, the transition function $\mathscr{P}$, the reward function $\mathscr{R}$, and the discount factor $\gamma$. At each timestep $t$, the agent finds itself in a state $s_t \in \mathscr{S}$ and chooses an action $a_t \in \mathscr{A}(s_t)$. The agent then receives a reward $R_{t+1}$ for the state-action pair $\mathscr{R}: \mathscr{S} \times \mathscr{A} \rightarrow \mathbb{R}$ and the environment transitions to the next state $s_{t+1}$ as by the transition function $\mathscr{P}: \mathscr{S} \times \mathscr{A} \times \mathscr{S} \rightarrow [0,1]$. The goal is to maximize the discounted sequence of rewards observed after timestep $t$, also called the expected return:
\begin{equation} \label{eq7}
	G_t = R_{t+1} + \gamma R_{t+2} + \gamma^2 R_{t+3} + ... = \sum_{k=t+1}^{T}\gamma^{k-t-1} R_{k},
\end{equation}
with $\gamma \in [0,1]$ \cite{a66_sutton2018reinforcement}. For this work, we define state space, action space, and reward function as follows:

\textit{State space}: A state $s_t \in \mathscr{S}$ is defined as $s_t = (c_{w,t}, P_{RE,t}, P_{D,t}, SOC_t, sin(h), cos(h), \mathbbm{1}(workday))$ and conveys the current timestep's information regarding wholesale price, renewable energy production, power demand, and battery SOC. In addition, $sin(h), cos(h), \mathbbm{1}(workday)$ are time-related features to help the agent recognize patterns, for example in the factory's power demand. $sin(h)$ and $cos(h)$ are a cyclical encoding for the hour of the day. $\mathbbm{1}(workday)$ is a binary feature that returns one if the current day is a workday for the factory and 0 otherwise.

\textit{Action space}: The action space is continuous and one-dimensional with $a_t = P_{B,t}$ and $a_t \in [P^{min}_{B}, P^{max}_{B}]$. We prevent the agent from violating the constraints \ref{eq4} - \ref{eq6} by applying the following security layer to obtain a corrected action, $a_{c,t}$ before executing the action in the environment:
\begin{equation} \label{eq9}
	\scalebox{0.95}[1]{$
	a_c,t = 
	\begin{cases}
		min(a_t, P^{max}_{B}, \frac{(SOC_{t}-SOC^{min})CaV}{\Delta t}) & , a_t \geq 0\\
		max(a_t, P^{min}_{B}, -P_{RE,t},\frac{(SOC_{t}-SOC^{max})CaV}{\Delta t}) & , a_t < 0
	\end{cases}
	$}
\end{equation}

This serves as an action mask correcting invalid actions to the nearest possible valid action.

\textit{Reward function}: The immediate reward, $R_t$, consists of two terms:
\begin{equation} \label{eq10}
	R_t = -C_{G,t} - \omega \mathbbm{1}(a_{c,t}, a_t),
\end{equation}
where $C_{G,t}$ is the cost of grid interaction as introduced by eq. \ref{eq1b} and $\omega \mathbbm{1}(a_{c,t}, a_t)$ is a penalty term with weight $\omega$ that is added only if $a_{c,t} \neq a_t$, i.e. when action correction is required. Both terms receive a negative sign to turn the cost minimization task into a reward maximization task. The penalty term is only used inside the control algorithm. The accumulated rewards reported in section \ref{casestudy} are computed as $R = -\sum_{t=0}^{T}C_{G,t}$.

We apply SAC to this RL framework. The algorithm has been introduced by Haarnoja et al. \cite{a203_haarnoja2018sac1} and improved by the same authors in \cite{a204_haarnoja2018sac2}. We implement the improved version which includes entropy regularization by learning the temperature parameter. Proposed as off-policy algorithm for continuous action spaces, SAC is a natural choice for the battery dispatch environment and our objective to utilize demonstration data. Algorithm \ref{algo1} shows our SAC variant in pseudocode, with focus on the modification we have made to gather demonstration data and incorporate it into the learning process. These modifications are:
\begin{itemize}
	\item Defining a deterministic, rule-based policy, $\pi_r$, that controls the environment through an if-then-else statement. For example, the rewards of the environment introduced in section \ref{problem} are greatly dependent on the current wholesale price, $c_{w,t}$. A straightforward rule could thus be:
	\begin{equation} \label{eq11}
		\pi_{r} \coloneqq a(s_t)= 
		\begin{cases}
			P^{max}_{B} & , c_{w,t} > \overline{c_w}\\
			P^{min}_{B} & , c_{w,t} \leq \overline{c_w}
		\end{cases}
		\hspace*{5pt}\forall t \in T
	\end{equation}
	which discharges the battery at $P^{max}_{B}$ whenever the current wholesale price $c_{w,t}$ is greater than the mean wholesale price $\overline{c_w}$, and charges at $P^{min}_{B}$ otherwise. 
	Designing such rules requires some domain knowledge and applying them to control the environment produces imperfect demonstration data. However, as we will show in the conducted case study, SAC benefits even from highly suboptimal rules and reliably learns policies that surpass them.
	\item Initializing a second replay buffer, $\mathcal{D}_{demo}$, and populating it with the transitions obtained from executing $\pi_r$ in the environment. In our experiments with episodes of 8760 timesteps, we conduct a single episode to collect demo data (see lines 2-6).
	\item Composing a joint batch from demonstrations and experiences by sampling from both replay buffers during training (see lines 14-18). The agent interacts with the environment using its own policy and stores the observed experiences in the replay buffer $\mathcal{D}_{exp}$, just as in the original SAC algorithm. To update network weights and temperature, a batch of size $\mathcal{B}$ is set together by uniformly sampling from both replay buffers according to ratio $\rho$ (see line 9). $\rho$ is designed to initially favor demonstration data and then shift towards the agent’s own experiences as training progresses. We use a linearly decaying $\rho$ and compute its value as $\rho = \frac{\mathcal{E}-e}{\mathcal{E}}$
	using the current episode, $e$, and the total number of episodes, $\mathcal{E}$.
\end{itemize}

The above modifications are easy to implement and insignificant regarding the added computational time. For the remainder of this paper, we will refer to algorithm \ref{algo1} as SAC from Demonstrations (SACfD).

\begin{algorithm}
	\caption{SAC from rule-based demonstrations}
	\label{algo1}
	\begin{algorithmic}[1]				
		\State \textbf{Initialize:} network parameters, rule-based policy $\pi_r$, empty replay buffer for demonstrations $\mathcal{D}_{demo}$, empty replay buffer for the agent's experiences $\mathcal{D}_{exp}$, size of minibatch $\mathcal{B}$
		\For{each step $t$ of one episode}:
		\State Obtain $a_t$ from $\pi_r(.|s_t)$
		\State Correct $a_t$ with the security layer
		\State Execute $a_t$ in environment
		\State $\mathcal{D}_{demo} \leftarrow \mathcal{D}_{demo}$ $\cup$ $\{s_t, a_t, r_{t+1}, s_{t+1}\}$
		\EndFor
		\For{each episode}:
		\State Obtain sampling ratio $\rho$
		\For{each step $t$}:
		\State Sample $a_t$ from $\pi(.|s_t)$
		\State Correct $a_t$ with the security layer
		\State Execute $a_t$ in environment
		\State $\mathcal{D}_{exp} \leftarrow \mathcal{D}_{exp}$ $\cup$ $\{s_t, a_t, r_{t+1}, s_{t+1}\}$
		\State $d_{demo} \leftarrow $ sample $\mathcal{B} \times \rho$ transitions from $\mathcal{D}_{demo}$
		\State $d_{exp} \leftarrow $ sample $\mathcal{B} \times (1-\rho)$ transitions from $\mathcal{D}_{exp}$
		\State $d \leftarrow d_{demo} \cup d_{exp}$
		\State \parbox[t]{313pt}{%
			Perform gradient step on all network weights\\ and temperature using $d$ \strut}
		\EndFor
		\EndFor	
	\end{algorithmic}
\end{algorithm}

\section{Case Study}
\label{casestudy}
\subsection{Parameter Setting}
To test the proposed method, we construct a case study of a hybrid power plant operating in Alberta, Canada. The location is assumed to be 49.05\degree N, 112.76\degree W and our experiments are based on hourly open-source data for the year 2018. The components of the plant are parameterized as listed in table \ref{tbl:params}. We design the case study to allow charge cycles across multiple timesteps, for example through the component sizes and the difference between purchasing and selling prices for power. Wholesale market prices for Alberta are taken from the Alberta Electric System Operator \cite{aeso}. The mean wholesale price for 2018 is 50 C\$/MWh. We use this mean value to generate demonstration data following eq. \ref{eq11} and call the corresponding rule-based policy $\pi_{r,50}$. Power profiles for renewable energies are downloaded from Renewables.ninja, an online tool simulating hourly wind and solar power output. Details of the tool are explained in \cite{a205_PFENNINGER2016_PV} for PV and in \cite{a206_STAFFELL2016_wind} for wind. For PV, we assume a system loss of 10\%, and fixed, southwards facing panels with a tilt angle of 53 degrees. For wind, we select five Siemens Gamesa SG 4.5 145 turbines and assume a hub height of 107.5m. The chosen location has high renewable resources, with capacity factors of almost 17\% for PV and 50.8\% for wind in the selected configuration for 2018.

\begin{figure}
	\begin{center}
		\includegraphics[width=8.4cm]{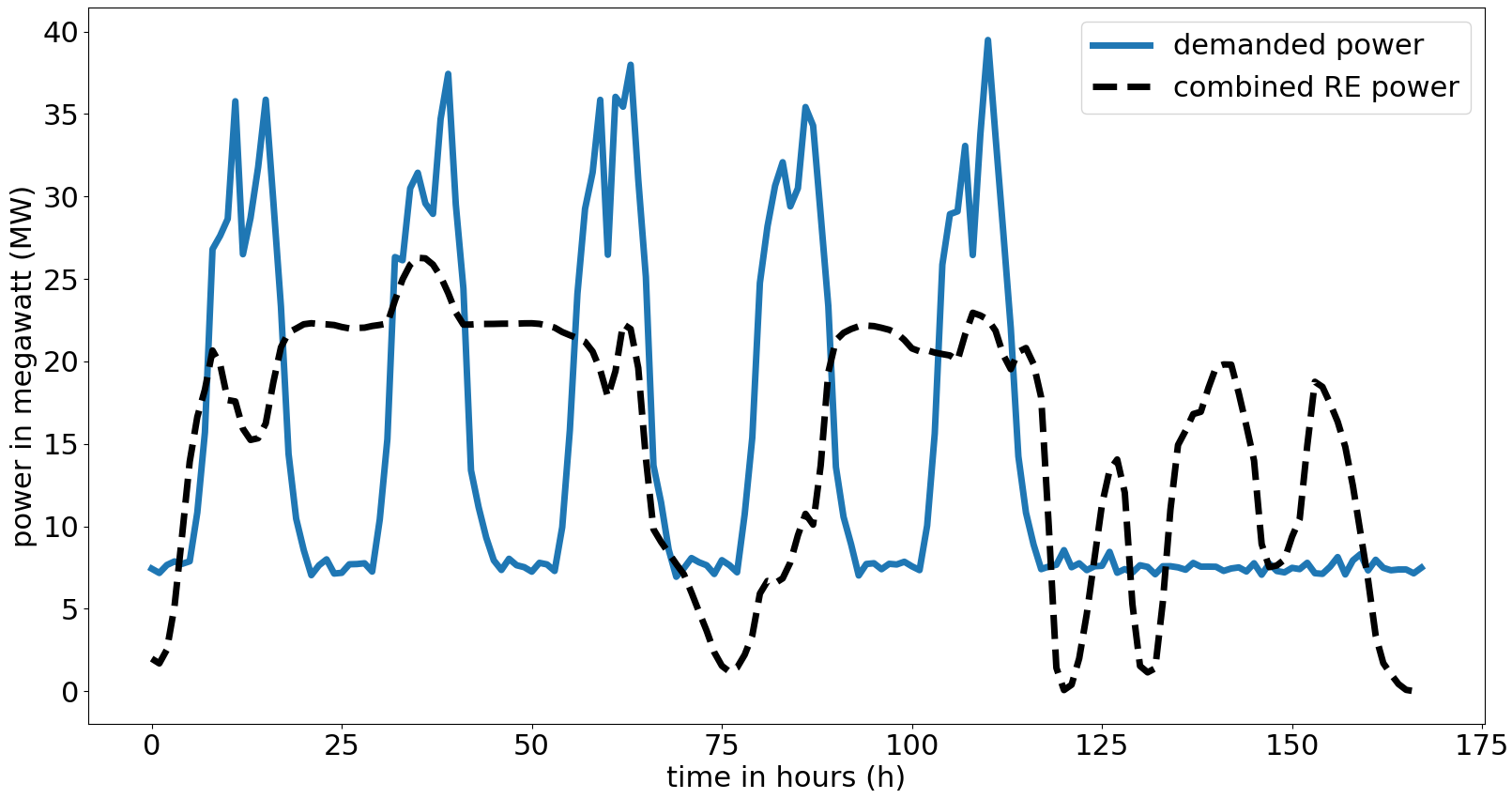} 
		\caption{Demand and RE power production during a random week in 2018.} 
		\label{fig:demand}
	\end{center}
\end{figure}

The power demand in the simulated microgrid stems from a factory, more specifically a manufacturing plant with a single shift operation. Due to the lack of suitable public data, we generate the load profile synthetically. The profile is obtained by taking a weekly base template with demand peaks before and after noon, and then adding random Gaussian noise to entire weeks as well as single hours. We further increase the demand during hot and cold periods and assume no plant operation on weekends and holidays. The resulting profile has significant demand fluctuations between days and nights as well as between workdays and days off. Figure \ref{fig:demand} shows an example of the demand for a random week in 2018 and compares it to the concurrent renewable power production. 

\begin{table}[]
	\begin{center}
	\caption{Parameters of the case study}
	\label{tbl:params}
	\begin{tabular}{l|l}
		\textbf{Parameter}  & \textbf{Value}       \\ \hline
		battery capacity, $CaV$        & 100 MWh     \\
		$SOC^{min}$/ $SOC^{max}$ & 0.2 / 0.8  \\
		$P^{min}_{B}$ / $P^{max}_{B}$  & -20 MW / 20 MW \\
		charge/ discharge efficiency, $\eta$        & 0.92 / $\frac{1}{0.92}$\\
		self-discharge, $\sigma$      & 0           \\
		$P_{wind, installed}$      & 22.5 MW     \\
		$P_{PV, installed}$        & 5 MW        \\
		aux. cost factor, $c_a$       & C\$ 10     
	\end{tabular}
	\end{center}
\end{table}

\subsection{Results and Discussion}

\begin{figure*}
	\begin{center}
		\includegraphics[width=\textwidth]{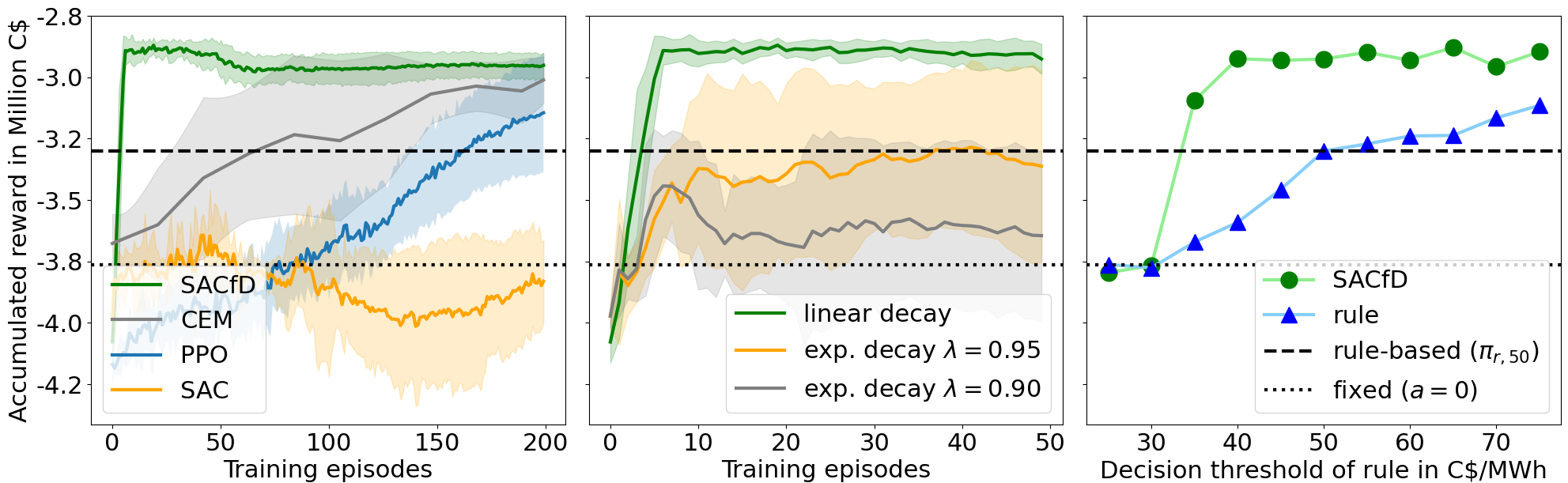} 
		\caption{Results of the conducted experiments. All experiments were repeated over five independent runs. The shaded area corresponds to +/- one standard deviation. (LEFT) Comparing the accumulated episodic reward of SACfD to three benchmark algorithms for long training periods of 200 episodes. (MIDDLE) Comparing SACfD's linearly decaying sampling ratio between demonstration and experience buffer to exponential decay rates. (RIGHT) Opposing the performance for rules with different decision thresholds to the performance of SACfD when trained with demonstration data from these rules.} 
		\label{fig:results}
	\end{center}
\end{figure*}

We first compare the performance of our proposed method to popular benchmarking algorithms. These are regular SAC \cite{a204_haarnoja2018sac2} and PPO \cite{a121_schulmanPPO} as deep actor-critic alternatives, as well as the cross-entropy method (CEM) \cite{a189_rubinstein1999cross} as heuristic alternative. We conduct informal hyperparameter tuning for each algorithm and compare the best configurations found over 200 episodes of training and five independent runs. The results are visualised in the left plot of figure \ref{fig:results}. SACfD is trained with rule-based demonstrations from the policy $\pi_{r,50}$. The dashed horizontal line shows the total episodic reward obtained when this policy controls the environment. The dotted horizontal line represents an agent with fixed action selection ($a_t = 0, \forall t\in T$).

SACfD outperforms the benchmark models in both sample efficiency and accumulated rewards. Especially the initial performance is remarkable as SACfD requires only 7 episodes of training until it peaks. The slight degrade in performance after around 50 episodes might be due to a suboptimal hyperparameter choice. Despite significant tuning efforts, both PPO and SAC struggle on this task. PPO requires long training time, and its performance is not stable across the five independent runs, leading to on average lower rewards. SAC fails to learn a useful policy and performs comparable to fixed action selection, a phenomenon that we frequently observed during hyperparameter tuning for both SAC and PPO. The performance for SAC shown in figure \ref{fig:results} (left) was obtained using the exact same parameters as for SACfD. This proves the considerable effect that the demonstration data has on the learning progress. The CEM learns better policies than PPO and SAC but still requires long training and has a large variance until late in training.

Next, we assess if exponentially decaying the sampling ratio $\rho$ can further improve initial training. Figure \ref{fig:results} (middle) shows 50 episodes of training with different decay rates. For two of the experiments, $\rho$ is computed as $\rho=\lambda^e$, with $e$ being the current episode. $\lambda=0.9$ represents a faster decay compared to linear, while lambda $\lambda=0.95$ decays slower than linear over 50 episodes. However, both variants perform worse than the linear decay. $\lambda=0.95$ struggles to outperform the rule used to collect demonstration data, while $\lambda=0.9$ causes the performance to degrade after some initial progress, as the share of demonstration data decreases quickly. We observe the same behavior when sampling uniformly from a single replay buffer storing both experiences and demonstrations. When ensuring sufficiently large buffers to avoid overwriting demonstrations, this effectively corresponds to setting $\rho=1/(e+1)$. Therefore, we conclude that the linearly decaying sampling rate offers a better balance between data from both replay buffers at the different stages of training.

The last experiment is to evaluate the robustness of SACfD to imperfect demonstration data. For this purpose, we vary the decision threshold of the if-then-else statements used to collect the demonstration data. While the rule-based policy $\pi_{r,50}$ was based on a threshold of 50 C\$/MWh, the mean wholesale price, we now experiment with price-based decision threshold from 25 – 75 C\$. In figure \ref{fig:results} (right) we compare the rewards obtained by the rules themselves (blue triangle) to the rewards obtained by SACfD after 50 episodes of training when the demonstration data stems from this rule (green circle). The results show that rules performing similar to fixed action selection (25 and 30 C\$/MWh thresholds) do not produce useful demonstration data. However, rules only slightly better than this can boost SACfD performance and allow the algorithm to reliably outperform the rules. The threshold of 35 C\$/MWh marks the transition between both. Interestingly, further improving the quality of the rule does not help to improve the final performance of SACfD. Once the rule-based demonstration data is of sufficient quality to initiate training, the reinforcement learning algorithm is able to adopt and continue training until convergence. 

\section{Conclusion}
We propose an approach to improve battery dispatch with RL through self-generated, imperfect demonstrations that are generated before training via simple if-then-else statements. This makes the approach applicable to tasks where no expert demonstrations are available. The results with SAC on the case study show that sample efficiency and accumulated rewards dramatically improve. Despite the simplicity of the proposed method, the use of demonstrations is the key factor to obtain well-performing and stable policies. The results further indicate that our method is robust to different rules, as long as the demonstration data can give the agent a nudge into the right direction during initial training.

A key challenge in the field of battery dispatch is the wide range of applications and the lack of a suitable benchmarking environment \cite{a194_subramanya9777914}.
Further research is required to assess the applicability of the approach to both other RL algorithms and other battery dispatch experiments.
A natural extension of the approach would be experiments with prioritized instead of uniform data sampling, as well as alternatives to price-based rules for the generation of demonstration data.
\section*{Acknowledgment}

This work is supported by the McGill Engineering Doctoral Award (MEDA) and MITACS grant number IT13369. The authors would like to thank Siemens Energy in Montreal for the support.



\bibliographystyle{IEEEtran}
\bibliography{references}
%



\end{document}